# Type-Aware Retrieval-Augmented Generation with Dependency Closure for Solver-Executable Industrial Optimization Modeling

Yuanjian Zhong[a], Rui Huang[a], Mengyao Wang[a], Zixin Guo[a], Yi-Chang Li[a,*], Mengmeng Yu[a,*], Zhong Jin[a]

[a]*School of Petroleum, China University of Petroleum (Beijing) at Karamay, Karamay, 834000, China,*

**Abstract**

Automated industrial optimization modeling requires reliable translation of natural-language requirements into solver-executable code. However, large language models often generate non-compilable models due to missing declarations, type inconsistencies, and incomplete dependency contexts. We propose a type-aware retrieval-augmented generation (RAG) method that enforces modeling entity types and minimal dependency closure to ensure executability. Unlike existing RAG approaches that index unstructured text, our method constructs a domain-specific typed knowledge base by parsing heterogeneous sources, such as academic papers and solver code, into typed units and encoding their mathematical dependencies in a knowledge graph. Given a natural-language instruction, it performs hybrid retrieval and computes a minimal dependency-closed context, the smallest set of typed symbols required for solver-executable code, via dependency propagation over the graph. We validate the method on two constraint-intensive industrial cases: demand response optimization in battery production and flexible job shop scheduling. In the first case, our method generates an executable model incorporating demand-response incentives and load-reduction constraints, achieving peak shaving while preserving profitability; conventional RAG baselines fail. In the second case, it consistently produces compilable models that reach known optimal solutions, demonstrating robust cross-domain generalization; baselines fail entirely. Ablation studies confirm that enforcing type-aware dependency closure is essential for avoiding structural hallucinations and ensuring executability, addressing a critical barrier to deploying large language models in complex engineering optimization tasks.

*Correspondence: liyic@cupk.edu.cn (YC.L.); ymm929@gmail.com (M.Y.).

*Keywords:* Retrieval-augmented generation; Type-aware knowledge representation; Dependency closure; Industrial optimization modeling

## 1. Introduction

Solver-executable industrial optimization modeling refers to the process of formulating engineering requirements, operational constraints, and decision objectives into mathematical programs that can be directly processed and solved by optimization solvers such as LINGO, Gurobi, or CPLEX to generate optimal or near-optimal decisions in manufacturing, energy, logistics, and other industrial systems. This modeling paradigm underpins a broad spectrum of intelligent industrial applications, ranging from distributed flexible assembly scheduling and embodied-AI-enabled production to carbon-aware energy scheduling and profit-driven low-carbon manufacturing (Jiang et al., 2026; Li et al., 2025; Liang et al., 2025; Xu et al., 2025). In such settings, optimization plays an indispensable role in balancing production efficiency, energy cost, and operational feasibility under strict physical and managerial constraints. As manufacturing systems become increasingly interconnected, the strong coupling across processes and tight temporal dependencies further amplify modeling complexity and elevate the cost of modeling errors (Lei et al., 2022; Yang et al., 2025b). These trends underscore the growing need to automate the translation from engineering requirements to executable optimization models.

Despite this pressing need, transforming requirements into formal optimization models remains challenging in constraint-intensive environments. Expert-driven modeling is often time-consuming and iterative; complex scheduling problems, such as flexible job shop scheduling, frequently require extensive manual debugging to ensure that variables, constraints, and objectives are complete, consistent, and solver-compliant (Lei et al., 2022; Yang et al., 2025b). Although recent industrial AI methods can improve downstream performance once a correct model is available, the upstream step of constructing the model itself continues to limit deployment speed and scalability (Cheng et al., 2024; Ma et al., 2024; Salazar et al., 2024). Large language models (LLMs) have shown considerable promise for automated modeling due to their ability to process natural language inputs and generate corresponding code (Jiang et al., 2025; Tang et al., 2025; Yang et al., 2025a). However, without explicit grounding in domain constraints, LLMs often produce incorrect references or structural hallucinations, rendering the resulting optimization models infeasible or unsolvable (Azamfirei et al., 2023; Zhang et al., 2025).

Retrieval-augmented generation (RAG) has been introduced to ground LLM outputs by supplying externally retrieved evidence (Lewis et al., 2020; Xiong et al., 2024), and recent variants, including GraphRAG and Self-RAG, aim to improve faithfulness and robustness (Asai et al.; Peng et al., 2025). For optimization-oriented tasks, OPT2CODE demonstrates that a RAG pipeline can translate problem descriptions into solver-executable code by retrieving documentation and applying selection mechanisms (Ahmed and Choudhury, 2025). Related work further explores structured retrieval beyond isolated chunks (e.g., KG$^2$RAG) and type-aware decomposition strategies (e.g., Typed-RAG) to better control retrieval and synthesis (Lee et al., 2025; Zhu et al., 2025), while operations-research surveys highlight automatic modeling as an emerging direction for LLM-enabled optimization (Wang and Li, 2025). Despite these advances, a critical gap remains for solver-executable optimization modeling: retrieved contexts are typically untyped text fragments or generic relations, which fail to guarantee the strict type consistency (e.g., variable vs. parameter roles) and dependency closure required to assemble executable optimization code (Ahmed and Choudhury, 2025; Lee et al., 2025; Lewis et al., 2020; Xiong et al., 2024; Zhang et al., 2025; Zhu et al., 2025). In solver-based modeling, every constraint must reference declared symbols with correct indexing and domains, and each modeling action must be supported by a minimal, dependency-closed set of upstream definitions. To clarify how these limitations differ across representative approaches, Table 1 provides a focused comparison of related work with respect to knowledge sources, retrieval granularity, and solver-executable support.

To address these limitations, we introduce a knowledge-grounded retrieval generation paradigm that explicitly models the typed dependencies inherent in optimization problems. Unlike conventional RAG which treats retrieved passages as isolated text chunks, our approach constructs a heterogeneous knowledge graph from both academic literature and executable solver code, capturing entities such as variables, parameters, constraints, and objectives along with their mathematical relations. Given a natural language request, the system first identifies relevant seed entities via hybrid semantic and structural retrieval, then automatically expands the context by traversing dependency edges to form a minimal closure that includes all symbols required for executability. This ensures that the generated code is both syntactically complete and semantically grounded, eliminating the missing declarations and type mismatches prevalent in prior work. We evaluate the proposed method on two structurally distinct industrial optimization tasks: demand response in battery manufacturing (Li et al., 2025; Lu et al., 2020) and flexible job shop scheduling (Lei et al., 2022; Yang et al., 2025b). Experimental results show that our method consistently yields compilable, solvable models that achieve optimal or near-optimal solutions, while baseline RAG approaches fail to produce executable code in all test instances. Furthermore, ablation studies reveal that both the integration of heterogeneous sources and the enforcement of dependency closure are critical to achieving this reliability, confirming that the proposed design effectively bridges the gap between high-level requirements and solver-ready implementations.

**Table 1**
Comparison of related RAG approaches

| Aspect | **This Paper** | OPT2CODE | KG$^2$RAG | Typed-RAG | Self-RAG |
| --- | --- | --- | --- | --- | --- |
| Primary Goal | Industrial Optimization Modeling | LP Code Generation | Retrieval coherence in QA | Type-based decomposition | Faithfulness critique |
| Knowledge Sources | Papers + Solver Code | PDF Manuals Only | Text chunks + KG relations | Text chunks + sub-queries | Text retrieval + self-critique |
| Typed Entities | Yes (Structured Entities) | No (Direct NL-to-Code) | Not domain-typed | Query types only | No (General) |
| Dependency Graph | Yes (Modeling Graph) | No (Linear Pipeline) | Yes (Fact-based KG) | No | No |
| Context Logic | Dependency Closure | Local Text Window | Partial evidence expansion | No | No |
| Paper–Code Alignment | Yes (Concept-to-Construct) | N/A | N/A | N/A | N/A |
| Validation | Solver Execution Loop | LLM Static Judges | Not solver-based | Not solver-based | Textual critique |

Our main contributions are as follows:

- We propose a type-aware retrieval-augmented generation method with dependency closure, which exposes the critical vulnerability of structural hallucinations in LLM-based industrial optimization modeling by enforcing symbolic completeness through typed knowledge graph construction.

- We design a two-stage retrieval mechanism that combines heterogeneous source parsing (academic papers and solver code) with minimal dependency closure computation, ensuring that generated models are syntactically complete and free of missing declarations or type inconsistencies.

- Our empirical results demonstrate that the method consistently produces compilable and solvable optimization models across two distinct industrial cases (battery production demand response and flexible job shop scheduling), where conventional RAG baselines fail entirely, validating its reliability and cross-domain robustness.

The remainder of this paper is organized as follows: Section 2 introduces the methodology, Section 3 presents the case studies and results, and Section 4 concludes the paper.

## 2. Knowledge Graph-Enhanced Type-Aware Retrieval with Dependency Closure for Reliable Industrial Optimization Modeling

Fig. 1 presents the proposed methodology for automated generation of solver-executable optimization models with a strong emphasis on reliability, which is the property that every generated model is guaranteed to be compilable, solvable, and free of structural hallucinations. The exposition focuses on the core innovations, including typed knowledge graph construction from heterogeneous sources, hybrid retrieval, and minimal dependency closure computation, which together enforce symbolic completeness and eliminate the missing declarations and type inconsistencies that plague conventional approaches.

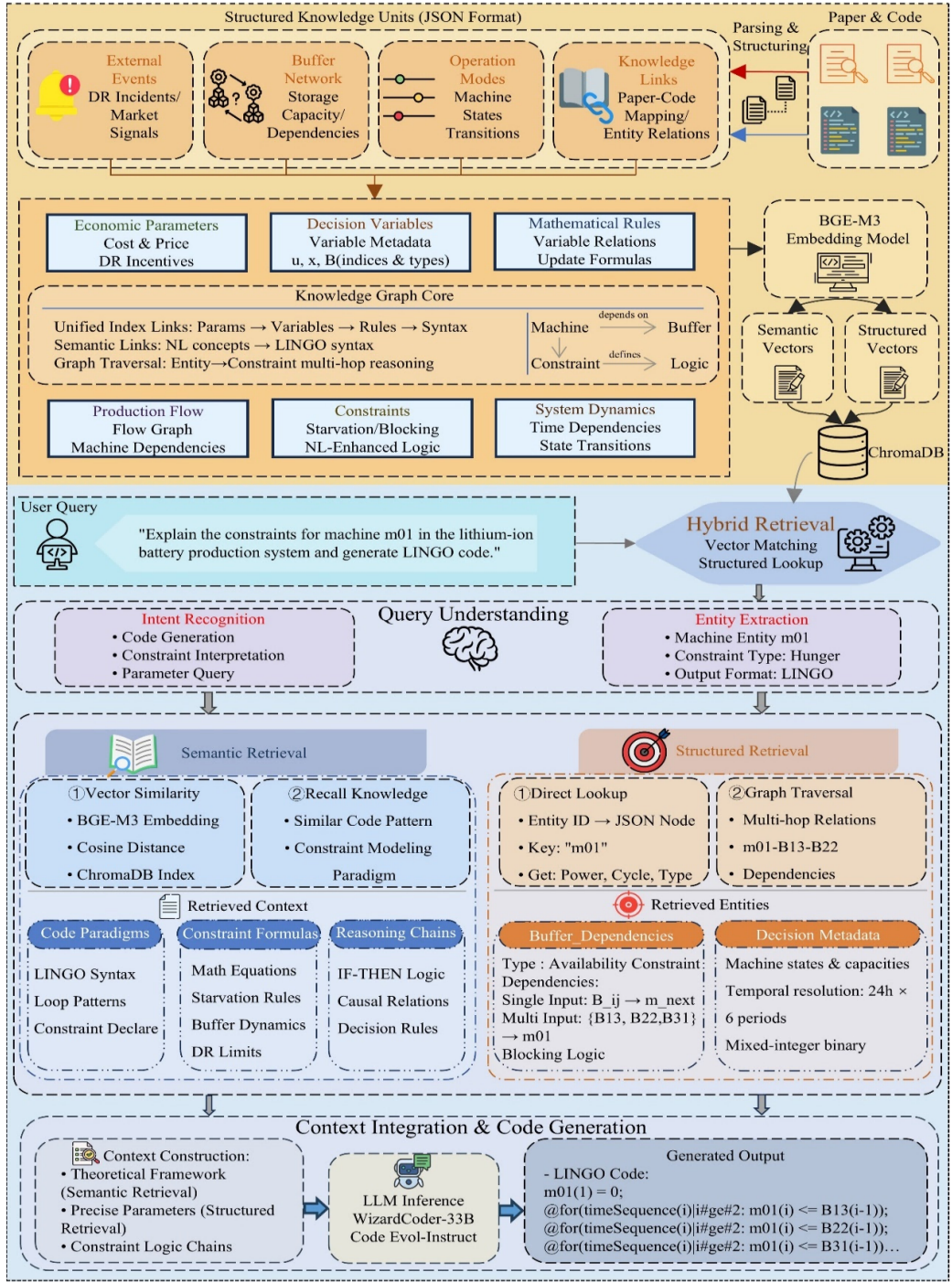

**Fig. 1.** Proposed Methodology for Reliable Industrial Optimization Modeling

## 2.1. Problem Analysis and Design Rationale

Industrial optimization models require symbolic completeness: every variable, parameter, and index set used in constraints and objectives must be explicitly declared, correctly typed, and properly referenced. Generic RAG methods retrieve semantically similar text fragments, but they cannot guarantee that the retrieved fragments together form a dependency-closed set of symbols. As a result, language models augmented with such RAG often generate code with missing declarations or type inconsistencies, which are structural hallucinations that render models non-executable. Achieving reliable automated modeling demands a paradigm shift from semantic relevance to dependency-aware completeness.

To overcome this, we identify three requirements absent in existing approaches, which form the foundation of a reliable methodology:

1. Explicit typing of modeling entities (variables, parameters, constraints, objectives, index sets) to prevent type mismatches.
2. Preservation of mathematical dependencies among those entities (e.g., a constraint depends on variables; a variable depends on index sets and parameters) to ensure all references can be resolved.
3. Retrieval of a minimal dependency-closed context for each generation action, which is the smallest set of upstream entities that makes the target well-defined and directly usable in solver code, guaranteeing that no necessary symbol is omitted.

These requirements drive the design of our methodology, whose primary goal is reliability in solver-executable model generation.

## 2.2. Overall Methodology

Fig. 1 illustrates the three-stage pipeline that operationalizes reliability:

1. Structured knowledge construction (top of Fig. 1). Academic papers and solver code are parsed and converted into typed knowledge units serialized in JSON. These units cover optimization-relevant categories (e.g., economic parameters, decision variables, mathematical rules, constraints, buffer-network relations, and optimization events), and they are connected by explicit links encoding both entity relations and paper–code mappings.

2. Hybrid retrieval driven by query understanding (middle of Fig. 1). Given a user request, the system performs query understanding, including intent recognition (e.g., constraint interpretation, code generation, parameter query) and entity extraction (e.g., machine identifiers, constraint type, output format), and then executes semantic retrieval over vector embeddings and structured retrieval over the knowledge graph.

3. Context integration and code generation (bottom of Fig. 1). Retrieved evidence is merged into a dependency-closed context package and provided to an LLM, which generates solver-ready code (LINGO in our implementation). A lightweight validation step can detect any residual issues and trigger corrective re-retrieval, further enhancing reliability.

The methodology guarantees that every generated model is symbol-complete and executable, eliminating the structural hallucinations that undermine reliability in prior approaches.

### 2.3. Constructing a Typed Knowledge Graph from Heterogeneous Sources
#### 2.3.1. Typed Entity Schema

We define a type system aligned with mixed-integer linear programming (MILP) modeling conventions. The core categories are:

- Decision variables –binary, integer, or continuous.

- Parameters –coefficients, capacities, prices.

- Index sets —time steps, machines, buffers, jobs.

- Constraints –linear or non-linear relations.

- Objective functions –to be maximized or minimized.

- Auxiliary rules –logical conditions, event triggers.

Each entity is stored with a unique identifier, type label, description, structured fields (e.g., domains, index sets, default values), and linkage fields for graph relations. The schema is extensible to new problem classes, maintaining reliability across domains.

#### 2.3.2. Parsing and Aligning Heterogeneous Sources
Two parsers extract knowledge:

- Solver code parser: An abstract-syntax-tree based parser processes LINGO (or Python/Gurobi) code to extract variable declarations, constraint definitions, objective expressions, and symbol references. This ensures that the knowledge graph faithfully represents executable constructs.

- Academic paper parser: A type-guided extraction procedure scans PDF papers for mathematical symbols and their contextual semantics (e.g., "machine power", "incentive price"), mapping them to the type schema. This captures the conceptual intent behind the symbols.

A paper-code alignment step creates semantic links between paper-derived concepts and their concrete realizations in solver code, enabling cross-modal traceability. Such alignment is essential for reliability: it bridges the gap between high-level requirements and low-level implementations, ensuring that generated code correctly reflects the intended mathematical model.

*2.3.3. Knowledge Graph Encoding with Dependency Relations*

All entities and alignments are organized into a directed graph $G = (V, E)$. Nodes $V$ represent typed entities. Edges $E$ are typed according to modeling roles:

- used_in: a variable or parameter used in a constraint or objective.

- depends_on: a constraint depending on a parameter or index set.

- aligns_to: a paper-concept aligned with a code entity.

This graph explicitly encodes the mathematical dependencies required for executable modeling. By capturing these relations, the graph becomes a reliable computational scaffold that supports dependency-aware retrieval and closure computation.

*2.3.4. Dual-Mode Indexing*

Entities' textual fields (descriptions, explanations) are embedded using a dense retriever (e.g., BGE-M3) and stored in a vector database for semantic retrieval. The graph itself is stored in a graph database (e.g., Neo4j) for structured queries and traversal. Dual-mode indexing allows the system to retrieve both conceptual background and exact typed definitions, a key factor in achieving reliability.

*2.4. Query-Driven Hybrid Retrieval*

Given a natural-language instruction, the system performs:

- Intent recognition and entity extraction: A lightweight parser determines the intent (e.g., add constraint, modify objective) and extracts domain entities (machine IDs, time windows, etc.). Accurate parsing is the first step toward reliable retrieval.

- Semantic retrieval: The query is embedded and matched against the vector index to obtain conceptual background (e.g., constraint explanations, modeling rationales). This provides the LLM with the necessary context to interpret the instruction correctly.

- Structured retrieval: Starting from identified seed nodes, the system traverses the knowledge graph along used_in and depends_on edges to collect related typed entities. This ensures that the retrieved set includes all symbols that could be required by the target.

The two retrieval streams are fused into a unified context package, balancing conceptual understanding with precise symbolic definitions, a combination that is essential for reliable code generation.

*2.5. Ensuring Executability via Minimal Dependency Closure: The Cornerstone of Reliability*

The central methodological contribution that guarantees reliability is the minimal dependency closure. This mechanism transforms the retrieved set of entities into a context that is both complete and minimal, eliminating any chance of missing declarations.

*2.5.1. Formal Graph Model*

Let $G = (V, E)$ be the knowledge graph with edge types used_in and depends_on considered executability-critical. A node $v$ is well-defined if all nodes reachable via these edges are included in the context.

*2.5.2. Computing the Minimal Dependency-Closed Context*

For a target entity $e \in V$ (e.g., a constraint to be generated), its dependency closure $C(e)$ is the smallest set satisfying:

- $e \in C(e)$;

- If $v \in C(e)$ and there exists an edge $(v \to u) \in E$ of type used_in or depends_on, then $u \in C(e)$;

- No other nodes are included.

$C(e)$ is computed by a breadth-first traversal restricted to the two edge types, using a visited set to handle cycles. The result is a minimal set containing exactly the symbols necessary to make $e$ well-defined, and excluding unrelated background. This minimality is crucial for reliability: it prevents the inclusion of extraneous symbols that could confuse the LLM or lead to unintended side effects.

*2.5.3. Context Packaging and Code Generation*

The final prompt sent to the language model combines:

- The typed entities in $C(e)$, each with its formal definition, domain, index sets, and solver-level symbol name;

- A textual description of the dependency subgraph induced by $C(e)$;

- The top-k semantic snippets retrieved earlier, providing explanatory context.

This package is fed to an LLM (e.g., WizardCoder-33B) with instructions to generate solver-ready LINGO code. The LLM's role is primarily translational; the critical reasoning about what to include has already been performed by the closure computation. By design, this process guarantees that the resulting code references only symbols that are defined within the context, ensuring reliability at the syntactic and semantic levels.

*2.5.4. Example*

Consider a query: "Add a load-reduction constraint for the demand-response event in the battery production case." The target is the constraint node representing the load-reduction inequality. Starting from this node, the traversal follows used_in edges to the involved variables (e.g., machine power variables) and depends_on edges to the relevant parameters (reference consumption, minimum reduction) and index sets (time steps). The closure $C(e)$ includes all symbols required to define the constraint, yielding a compilable LINGO model. Conventional RAG, lacking dependency awareness, would omit essential symbols and produce non-executable code, a failure of reliability that our methodology systematically avoids.

## 2.6. Reproducibility

To support reproducibility and facilitate community adoption, all components produced by the construction pipeline, including the core type schema, extracted knowledge units (in JSON), the constructed knowledge graph, and the embedding vectors, are archived in a public repository (DOI: 10.5281/zenodo.18491531).

## 3. Case Studies and Results

This section validates the proposed method through two industrial cases: demand response optimization in lithium-ion battery module production, which involves strongly coupled machine starvation and buffer blocking constraints with dynamically adjusted objective functions based on grid incentives, and the flexible job shop scheduling problem (FJSP), which tests cross-domain generalization due to its distinct variable types and constraint structures. Both cases follow a consistent experimental paradigm: a baseline optimization model is constructed and stored in the knowledge base, then modified via natural language instructions, with the generated code evaluated for compilability, solvability, and optimization quality. The experiments are divided into two parts: Sections 3.1 and 3.2 demonstrate the method's ability to generate executable code and obtain high-quality solutions, while Section 3.3 uses ablation studies and baseline comparisons to verify the contributions of heterogeneous knowledge source integration and type-aware retrieval. The complete experimental configuration is summarized in Table 2.

### 3.1. Demand Response Optimization in Lithium-Ion Battery Module Production

#### 3.1.1. Case Background and System Layout

This case is based on the lithium-ion battery module production system described in (Li et al., 2025; Lu et al., 2020). Fig. 2 illustrates the system layout, where each $m_{ij}$ represents the $j$-th machine in the $i$-th production line branch, and each $B_{ij}$ denotes the buffer that stores intermediate products from machine $m_{ij}$ (refer to (Li et al., 2025) for detailed machine operating parameters and constraints). As shown in the figure, three parallel upstream production lines converge into a single core assembly line. This converging configuration introduces coupled constraints through machine starvation and buffer blocking, making it a testbed for evaluating the proposed method.

**Table 2**
Experimental Environment and Parameter Configuration

| Category | Item | Configuration/Parameter |
|---|---|---|
| **Computing Platform** | Processor | Intel Core i9-14900K (24 cores) |
| | Memory | 128 GB |
| | Solver | LINGO 18.0 / python3.9 |
| **Core System Components** | LLM Model | wizardcoder:33b-v1.1 |
| | Embedding Model | BGE-M3 |
| | Vector Database | ChromaDB |
| **Knowledge Base** | Scale | 40 academic papers and their solver models archived in a public repository (DOI: 10.5281/zenodo.18491531) |

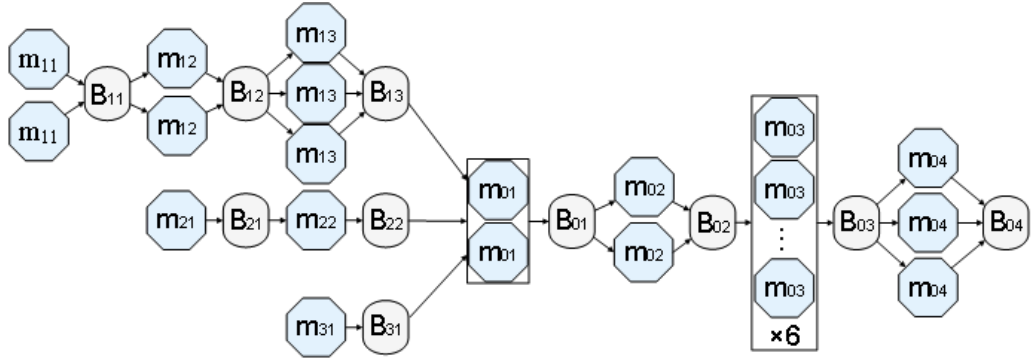

**Fig. 2.** Layout of the battery module assembly system.

*3.1.2. Optimization Model and Demand Response Scenario*

This study adopts incentive-based demand response (IBDR), which provides direct economic rewards for load reduction. The manufacturer's optimization goal is to maximize profit while meeting production requirements, formulated as a MILP problem.

The experimental scenario considers a 24-hour scheduling horizon divided into 144 time steps of 10 minutes each. During the 16th–17th hours (time steps 91–102), the grid operator issues an IBDR event requiring participants to reduce load by at least 10 kW in exchange for an incentive price of 0.54 \$/kWh. Using our proposed method, the baseline model's objective function is automatically modified to balance incentive revenue against potential production loss.

In the absence of an IBDR event, the baseline objective function is defined as:

$$\max \Phi = \sum_{k=1}^{n} r_k q_k - \sum_{k=1}^{m-1} c_k n_k - \sum_{\tau \in T} p(\tau) \sum_{i \in I, j \in J} \left[ P_{\text{on},ij} y_{ij}(\tau) + P_{\text{off},ij}(1 - y_{ij}(\tau)) \right] \quad (1)$$

where $\Phi$ is the total profit of the manufacturer (in monetary units, e.g., \$); $r_k$ and $q_k$ are the unit price (\$/unit) and quantity produced (units) of the $k$-th output product respectively; $c_k$ and $n_k$ are the unit cost (excluding electricity, \$/unit) and quantity consumed of the $k$-th input material respectively; $T$ is the set of all time slots in the scheduling horizon; $p(\tau)$ is the electricity price at time slot $\tau$ (\$/kWh); $I$ and $J$ are the sets of production line branch indices and machine indices within each branch respectively; $P_{\text{on},ij}$ and $P_{\text{off},ij}$ are the per-time-slot energy consumption (kWh) of machine $m_{ij}$ when it is active and in low-power mode respectively; and $y_{ij}(\tau)$ is a binary decision variable, where $y_{ij}(\tau) = 1$ if machine $m_{ij}$ is scheduled ON during time slot $\tau$, and 0 otherwise.

When the IBDR event occurs (time steps 91–102), the objective function is automatically modified by our method to:

$$\max \Phi = \sum_{k=1}^{n} r_k q_k - \sum_{k=1}^{m-1} c_k n_k - \sum_{\tau \in T} \left[ p(\tau) \sum_{i \in I, j \in J} \Lambda_{ij}(\tau) \right] + \Omega(T^*) \quad (2)$$

with auxiliary definitions:

$$\Lambda_{ij}(\tau) = P_{\text{on},ij}\, y_{ij}(\tau) + P_{\text{off},ij} \left(1 - y_{ij}(\tau)\right) \quad (3)$$

$$\Omega(T^*) = \lambda(T^*) \left[ \sum_{\tau \in T^*} B_{\text{ref}}(\tau) - \sum_{\tau \in T^*} \sum_{i \in I,\, j \in J} \Lambda_{ij}(\tau) \right] \quad (4)$$

and the load reduction constraint:

$$\sum_{\tau \in T^*} \left[ B_{\text{ref}}(\tau) - \sum_{i \in I,\, j \in J} \Lambda_{ij}(\tau) \right] \geq \Delta L_{\min}$$

$$T^* \in \{\text{RT-DR event hours}\}$$

where $\Lambda_{ij}(\tau)$ is the actual energy consumption of machine $m_{ij}$ during time slot $\tau$ (kWh); $T^*$ is the set of time slots during which the IBDR event is active (e.g., slots 91–102); $\Omega(T^*)$ and $\lambda(T^*)$ are the total incentive payment received for participating in the IBDR event (\$) and the incentive rate offered during the event (\$/kWh) respectively; $B_{\text{ref}}(\tau)$ is the customer baseline load for time slot $\tau$ (kWh), typically calculated from historical consumption data; and $\Delta L_{\min}$ is the minimum load reduction required to qualify for incentive payments (kWh).

The baseline LINGO objective (without IBDR) stored in the knowledge base is:

Listing 1: Baseline LINGO objective function (without IBDR)

```
max = B04(144) * BATTERY_PROFIT
    - @sum(timeSequence(I) :
        (m11(I) * m11Power + m12(I) * m12Power
        + ...
        + m04(I) * m04Power)
        * DayAheadPrice(@floor((I - 1)/6) + 1))/6;
```

This code snippet represents the profit maximization in LINGO syntax: the revenue from final products ($B04(144) * \text{BATTERY\_PROFIT}$) minus the total electricity cost, where $m11(I)$ etc. are binary decision variables for each machine at each time interval, $m11Power$ etc. are the corresponding power ratings, and the electricity price is averaged over six 10-minute slots per hour.

*3.1.3. Results and Validation*

After receiving the IBDR instruction, the proposed method executes the hybrid retrieval strategy. Semantic retrieval returns descriptions of the incentive mechanism from the literature, while structured retrieval locates the objective function entity and its dependent variable/parameter nodes. The dependency propagation mechanism collects the complete context along the knowledge graph, including machine power parameters, time-step indices, and price arrays.

The generated modified objective code is shown in Listing 2.

Listing 2: LINGO code automatically generated by our method for DR objective modification

```
! 1. DR Period Energy Consumption Calculation
```

```
hour16Consumption=
@sum(timeSequence(I)|(I#gt#6*15)#and#(I#le#6*16):
  (m11(I)*m11Power + ... + m04(I)*m04Power))/6;
hour17Consumption=
@sum(timeSequence(I)|(I#gt#6*16)#and#(I#le#6*17):
  (m11(I)*m11Power + ... + m04(I)*m04Power))/6;

! 2. Modified objective function
!    (with added incentive reward component)
max = B04(144)*5 - @sum(...)
    + (Bref16 - hour16Consumption) * 0.54
    + (Bref17 - hour17Consumption) * 0.54;

! 3. Load Reduction Constraint
(Bref16 - hour16Consumption) +
(Bref17 - hour17Consumption) >= 10;
```

The model, as automatically modified by our method, converges to the global optimum within 120 seconds. Fig. 3 illustrates the hourly energy consumption profile of the production line with and without the IBDR event, where the incentive rate is set at 0.54 \$/kWh during the 16th and 17th hours. The blue bars represent the baseline load without IBDR, while the red bars show the optimized load under the IBDR event. A significant reduction in load during the event window (hours 16–17) is observed, demonstrating the successful adjustment of production schedules to achieve load shedding. After the event, the load quickly recovers, reflecting the flexibility of the production scheduling.

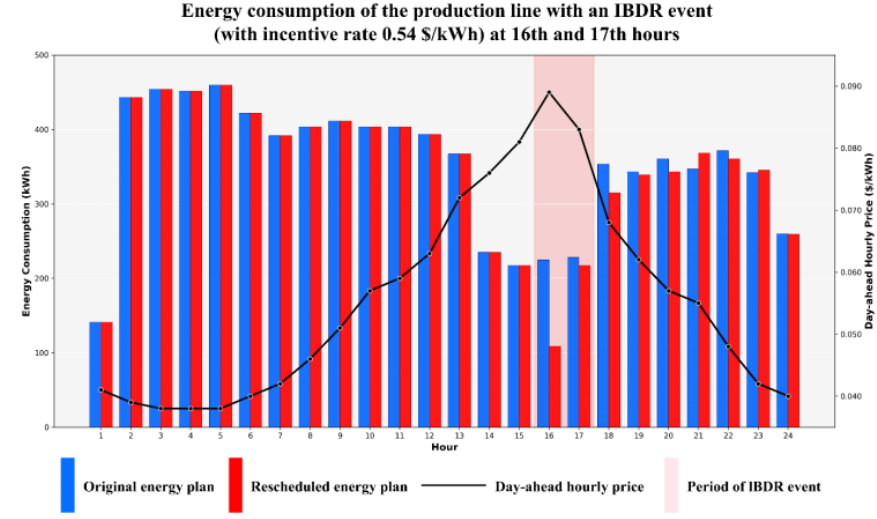

**Fig. 3.** Energy consumption of the production line with an IBDR event (with incentive rate 0.54 \$/kWh) at 16th and 17th hours

Fig. 4 shows the corresponding optimized on/off schedule for each machine across the 24-hour horizon (144 time slots of 10 minutes each). The horizontal axis represents the scheduling time, and the vertical axis lists all machines in the production line (as denoted in Fig. 2). Each cell in the grid indicates the machine's status during a specific time slot: blue cells represent the machine being scheduled "on" (active), while white cells represent "off" (low-power state). During the IBDR event window (hours 16–17, corresponding to time slots 91–102), a subset of machines are highlighted in red. These red cells denote machines that were originally planned to be "on" in the baseline schedule but are deliberately turned "off" by our method to curtail energy consumption and earn incentive rewards. The prevalence of red cells in this region directly explains the load reduction observed in Fig. 3. After the event concludes, machines are progressively turned back on to resume production, though the temporary halt results in a final production drop from 828 to 786 units (a 5.1% reduction). Nevertheless, profit increases from $2776.86 to $2780.51 (a 0.13% improvement) due to the incentive earnings,

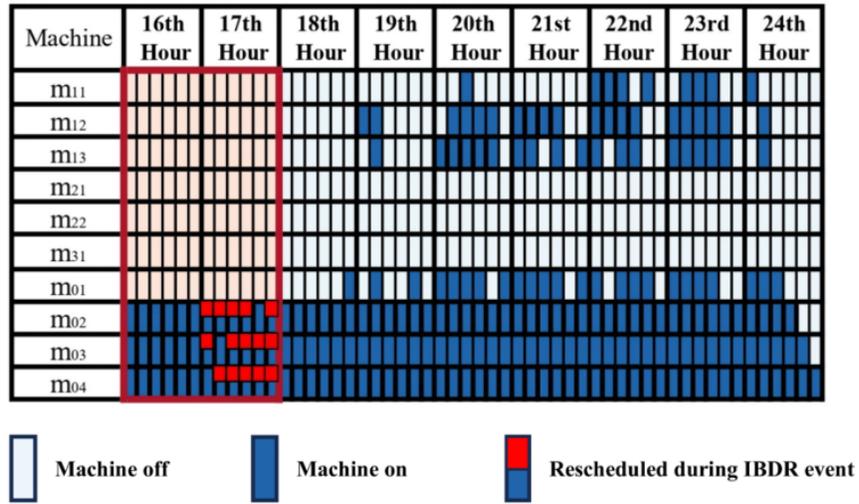

**Fig. 4.** Optimized Production Schedule and Machine Status During the Demand Response Event

demonstrating the method's ability to balance production revenue against IBDR incentives.

This detailed visualization of machine-level decisions provides insight into how the optimization algorithm dynamically adjusts operations to exploit incentive opportunities while respecting complex interdependencies (e.g., starvation and blocking constraints). The use of red cells to highlight rescheduled machines underscores the precision of our type-aware RAG method in generating executable, constraint-compliant schedules that would be infeasible with conventional RAG baselines.

*3.2. Production Scheduling Optimization for Flexible Job Shop Scheduling (FJSP)*

*3.2.1. Problem Description and Benchmark Instances*

To test cross-domain generalization, the FJSP is selected as the second case. FJSP differs significantly from battery production in variable types and constraint structures as shown in Table 3.

**Table 3**

Key Differences in Variable Types and Constraint Structures Between Lithium-ion Battery Production and FJSP

| Aspect | Lithium-ion Battery Production (Case 1) | Flexible Job Shop Scheduling (FJSP) (Case 2) |
| --- | --- | --- |
| **Decision Variables** | Machine ON/OFF states (binary) & Buffer inventory levels (integer) | Machine assignment for operations (binary) & Start times of operations (continuous) |
| **Core Constraints** | Machine starvation & Buffer material balance & Production line coupling | Precedence relations between operations & Machine capacity (one operation at a time) & Time window constraints |
| **Objective Function** | Maximize profit (product revenue - production cost - energy cost + IBDR incentive) | Minimize makespan (maximum completion time) |
| **Problem Nature** | Production flow control & energy management under IBDR events | Scheduling of jobs with flexible routing on machines |
| **Mathematical Structure** | MILP with strong spatial–temporal coupling | MILP with sequencing & assignment logic |

Experiments use Behnke benchmark instances with three variants:

- Variant 1 (baseline modeling): generate a MILP model minimizing makespan.

- Variant 2 (constraint extension): add machine unavailability windows.

- Variant 3 (semantic robustness): use alternative industrial terminology.

We use python 3.9 as the solver with a time limit of 120 seconds. Reported solving times are wall-clock times measured externally, which may slightly exceed the internal solver limit due to preprocessing, model loading, and output overhead.

*3.2.2. Experimental Variants and Results*

(1) Baseline modeling (Variant 1) ran 20 times independently. One run had an indexing error; the remaining 19 succeeded. This error stemmed from an off-by-one issue in loop boundary expressions, rather than from missing variable declarations or type inconsistencies, problems that our dependency closure mechanism is specifically designed to eliminate. This distinction highlights that while our method guarantees symbolic completeness (all variables, parameters, and sets are properly declared), it does not fully eliminate logical errors in complex indexing expressions, which depend on the LLM's translation fidelity and the intricacies of the target solver's syntax. Future work will integrate automated post-generation validation and self-correction mechanisms to detect and rectify such logical errors, further enhancing generation reliability. Nevertheless, the 95% success rate achieved on this challenging NP-hard problem substantially outperforms the 0% success rate of conventional RAG baselines, demonstrating the robustness and practical viability of our approach. Table 4 shows results on benchmark instances. All successfully generated models reach known optima under the configured 120-second solver time limit, with measured wall-clock solving times ranging from 115.1 to 125.6 seconds across multiple runs.

**Table 4**
Solving Performance of the Method-Generated FJSP Models on Benchmark Instances

| Instance | Size (Jobs×Machines) | Obtained Makespan (Optimal Solution) | Solving Time Range (seconds) |
| --- | --- | --- | --- |
| Behnke1.fjs | 10×20 | 110.0 | 118.9–120.4 |
| Behnke3.fjs | 10×20 | 120.0 | 119.0–121.1 |
| Behnke8.fjs | 20×20 | 262.0 | 115.1–115.4 |
| Behnke10.fjs | 20×20 | 283.0 | 115.3–115.9 |
| Behnke15.fjs | 50×20 | 1108.0 | 122.5–125.6 |

(2) Constraint extension (Variant 2) successfully introduced auxiliary binary variables and unavailability constraints; models remained compilable and solvable as below:

Listing 3: Generated Python snippet for machine unavailability window constraints
```
# Machine Unavailability Time Window Constraints (Method Generated Snippet)
for w_idx, (w_start, w_end) in enumerate(unavailability_windows[k]):
    p_ijk = operations_times[i, j, k]
    # Constraint 1: Operation completes before the maintenance window starts (if z
    =1)
    model += (s[i, j] + p_ijk <= w_start + largeM * (1 - z[i, j, k, w_idx]) +
    largeM * (1 - x[i, j, k]))
    # Constraint 2: Operation starts after the maintenance window ends (if z=0)
    model += (s[i, j] >= w_end - largeM * z[i, j, k, w_idx] - largeM * (1 - x[i, j
    , k]))
```

(3) Semantic robustness (Variant 3) correctly mapped alternative terminology via semantic links in the knowledge base, producing logically equivalent code.

*3.3. Ablation Study and Comparative Analysis*

*3.3.1. Ablation Study on Heterogeneous Knowledge Sources*

Using the instruction "explain the constraints of machine $m_{01}$ and generate corresponding LINGO code" from the lithium-ion battery case as an example, three knowledge source configurations are compared:

**Configuration 1: Academic papers only**

The generated code used an IF-THEN structure instead of the required @for loop, the variable name was a generic Buffer_1 (which should actually be $B_{13}$, $B_{22}$, and $B_{31}$), and it lacked time index dependencies. The generated code is as follows:

Listing 4: Generated code for m01 constraints using academic papers only
```
IF SUM([j IN Buffer_1], Level[j]) < CARD(Buffer_1) THEN
```

```
  "Machine doesn't run"
ELSE
  "Machine runs";
ENDIF...
```

**Configuration 2: Solver code only**

The generated constraint expressions lacked complete semantic descriptions, resulting in incomplete logic. The generated content is as follows:

Listing 5: Generated code snippet for $m_{01}$ using solver code only
```
DEFINE ENTITY m01 : TYPE = "machine", MAX_PRODUCTION = 100...
```

**Configuration 3: Heterogeneous structured knowledge base**

The generated constraint code used the correct LINGO syntax structure, included complete constraints for the three upstream buffers, and reflected time index dependencies: at each time step, $m_{01}$ can only operate if the inventory levels of buffers $B_{13}$, $B_{22}$, and $B_{31}$ at the previous time step meet the production requirements of $m_{01}$. The generated constraint code is as follows:

Listing 6: Generated LINGO code for $m_{01}$ constraints using heterogeneous structured knowledge base
```
m01(1) = 0;
@for(timeSequence(i)|i#ge#2: m01(i) <= B13(i-1));
@for(timeSequence(i)|i#ge#2: m01(i) <= B22(i-1));
@for(timeSequence(i)|i#ge#2: m01(i) <= B31(i-1))...
```

The comparative results demonstrate that a single knowledge source leads to either syntax deviations or insufficient semantic understanding. The heterogeneous knowledge base simultaneously provides semantic context and precise syntax, ensuring code correctness.

*3.3.2. Effectiveness of Type-Aware Retrieval with Dependency Closure*

On FJSP Variant 2, the proposed method succeeded in all 5 runs, while a conventional (non-type-aware) RAG baseline failed in all 5 runs due to undefined variables and indexing errors (Table 5).

*3.3.3. Discussion on Generality and Robustness*

The proposed method generates executable code in two structurally different industrial cases. Heterogeneous sources provide complementary seman-tics and syntax, while type-aware retrieval eliminates variable omission and type mismatch via dependency closure.

**Table 5**
Method Robustness Evaluation: Compilability and Constraint Integrity Comparison

| Method | Number of Runs | Model Compilable/ Solvable | Key Constraint Integrity | Remarks |
|---|---|---|---|---|
| **Proposed Method** | 5 | 5/5 | 5/5 | — |
| LLM+Regular RAG (Non-Type-Aware Baseline) | 5 | 0/5 | 0/5 | Failed due to various structural errors: missing imports, undefined variables, syntax errors, etc. |

Current validation covers only two cases, with 95% generation stability and medium-scale instances. Future work will extend to more scenarios (multi-objective, nonlinear), integrate automatic post-processing and multi-round self-correction, and test on large-scale and multi-day instances.

## 4. Conclusions

This work presents a type-aware RAG approach incorporating dependency closure to overcome the challenge of automatically generating solver-executable industrial optimization models from natural-language specifications. The core innovation lies in constructing a heterogeneous knowledge graph that explicitly encodes typed modeling entities (variables, parameters, constraints) and their mathematical dependencies, coupled with a minimal dependency closure mechanism that retrieves only the symbols necessary for executability. Validated on two distinct industrial tasks: demand response in battery production and flexible job shop scheduling, the method consistently produces compilable and solvable models that achieve optimal solutions, whereas conventional RAG baselines fail entirely. Ablation studies confirm that both heterogeneous source integration and dependency closure are essential for reliability. This work establishes a new paradigm for reliable LLM-based code generation in engineering domains, enabling non-experts to generate correct optimization models and significantly reducing manual debugging effort.

# References

Ahmed, T., Choudhury, S., 2025. Opt2code: A retrieval-augmented framework for solving linear programming problems. Natural Language Processing Journal , 100185.

Asai, A., Wu, Z., Wang, Y., Sil, A., Hajishirzi, H.S.R., . Learning to retrieve, generate, and critique through self-reflection. arxiv 2023. arXiv preprint arXiv:2310.11511 .

Azamfirei, R., Kudchadkar, S.R., Fackler, J., 2023. Large language models and the perils of their hallucinations. Critical Care 27, 120.

Cheng, H., Lu, T., Hao, R., Li, J., Ai, Q., 2024. Incentive-based demand response optimization method based on federated learning with a focus on user privacy protection. Applied Energy 358, 122570.

Jiang, G., Ma, Z., Zhang, L., Chen, J., 2025. Prompt engineering to inform large language model in automated building energy modeling. Energy 316, 134548.

Jiang, W., Deng, Q., Luo, Q., Zhang, J., Zhou, J., 2026. A knowledge-based memetic algorithm for integrated scheduling of equipment operation and spare parts manufacturing in distributed assembly flexible job shops. Engineering Applications of Artificial Intelligence 164, 113311.

Lee, D., Park, A., Lee, H., Nam, H., Maeng, Y., 2025. Typed-rag: Type-aware decomposition of non-factoid questions for retrieval-augmented generation, in: Proceedings of the 1st Joint Workshop on Large Language Models and Structure Modeling (XLLM 2025), pp. 129–152.

Lei, K., Guo, P., Zhao, W., Wang, Y., Qian, L., Meng, X., Tang, L., 2022. A multi-action deep reinforcement learning framework for flexible job-shop scheduling problem. Expert Systems with Applications 205, 117796.

Lewis, P., Perez, E., Piktus, A., Petroni, F., Karpukhin, V., Goyal, N., Küttler, H., Lewis, M., Yih, W.t., Rocktäschel, T., et al., 2020. Retrieval-augmented generation for knowledge-intensive nlp tasks. Advances in neural information processing systems 33, 9459–9474.

Li, Y.C., Wang, M., Huang, R., Chen, L., Wang, X., Xiong, X., Jiang, M., Cui, L., Jia, Z., Jin, Z., 2025. Profit-driven framework for low-

carbon manufacturing: Integrating green certificates, demand response, distributed generation and ccus. Energies 18, 6517.

Liang, Y., Pu, Y., Yu, M., Li, Y.C., Jiang, M., Cui, L., Xiong, X., Jin, Z., 2025. A dual carbon reward mechanism for electric vehicle charging scheduling based on multi-level stackelberg game. Energy , 138620.

Lu, R., Li, Y.C., Li, Y., Jiang, J., Ding, Y., 2020. Multi-agent deep reinforcement learning based demand response for discrete manufacturing systems energy management. Applied Energy 276, 115473.

Ma, S., Liu, H., Wang, N., Huang, L., Su, J., Zhao, T., 2024. Incentive-based integrated demand response with multi-energy time-varying carbon emission factors. Applied Energy 359, 122763.

Peng, B., Zhu, Y., Liu, Y., Bo, X., Shi, H., Hong, C., Zhang, Y., Tang, S., 2025. Graph retrieval-augmented generation: A survey. ACM Transactions on Information Systems 44, 1–52.

Salazar, E.J., Rosero, V., Gabrielski, J., Samper, M.E., 2024. Demand response model: A cooperative-competitive multi-agent reinforcement learning approach. Engineering Applications of Artificial Intelligence 133, 108273.

Tang, J., Chen, J., He, J., Chen, F., Lv, Z., Han, G., Liu, Z., Yang, H.H., Li, W., 2025. Towards general industrial intelligence: A survey of large models as a service in industrial iot. IEEE Communications Surveys & Tutorials .

Wang, Y., Li, K., 2025. Large language models in operations research: Methods, applications, and challenges. arXiv preprint arXiv:2509.18180 .

Xiong, G., Jin, Q., Lu, Z., Zhang, A., 2024. Benchmarking retrieval-augmented generation for medicine, in: Findings of the Association for Computational Linguistics: ACL 2024, pp. 6233–6251.

Xu, J., Sun, Q., Han, Q.L., Tang, Y., 2025. When embodied ai meets industry 5.0: Human-centered smart manufacturing. IEEE/CAA Journal of Automatica Sinica 12, 485–501.

Yang, X., Lin, C., Yang, Y., Wang, Q., Liu, H., Hua, H., Wu, W., 2025a.24

Large language model powered automated modeling and optimization of active distribution network dispatch problems. IEEE Transactions on Smart Grid .

Yang, Z., Li, X., Gao, L., Liu, Q., 2025b. A heterogeneous graph attention-enhanced deep reinforcement learning framework for flexible job shop scheduling problem with variable sublots. Engineering Applications of Artificial Intelligence 157, 111375.

Zhang, Z., Wang, C., Wang, Y., Shi, E., Ma, Y., Zhong, W., Chen, J., Mao, M., Zheng, Z., 2025. Llm hallucinations in practical code generation: Phenomena, mechanism, and mitigation. Proceedings of the ACM on Software Engineering 2, 481–503.

Zhu, X., Xie, Y., Liu, Y., Li, Y., Hu, W., 2025. Knowledge graph-guided retrieval augmented generation, in: Proceedings of the 2025 Conference of the Nations of the Americas Chapter of the Association for Computational Linguistics: Human Language Technologies (Volume 1: Long Papers), pp. 8912–8924.